\documentclass[aps,showpacs,nofootinbib]{revtex4}

\usepackage{graphicx}
\usepackage{graphics}
\usepackage{amssymb}
\usepackage{amsmath}
\usepackage{bm}

\newcommand{\sT}{{\scriptscriptstyle T}}

\newcommand{\be}{\begin{equation}}
\newcommand{\ee}{\end{equation}}
\newcommand{\bea}{\begin{eqnarray}}
\newcommand{\eea}{\end{eqnarray}}
\newcommand{\beal}{\begin{align}}
\newcommand{\eal}{\end{align}}
\newcommand{\bespl}{\begin{split}}
\newcommand{\espl}{\end{split}}
\newcommand{\nn}{\nonumber}
\newcommand{\nslash}{\kern 0.2 em n\kern -0.50em /}
\newcommand{\kslash}{\kern 0.2 em k\kern -0.45em /}
\newcommand{\pslash}{\kern 0.2 em p\kern -0.50em /}
\newcommand{\Sslash}{\kern 0.2 em S\kern -0.50em /}
\newcommand{\Pslash}{\kern 0.2 em P\kern -0.50em /}
\newcommand{\Rslash}{\kern 0.2 em R\kern -0.50em /}

\begin{document}

\title{Effective role of unpolarized nonvalence partons in Drell-Yan single spin 
asymmetries}

\author{Andrea Bianconi}
\email{andrea.bianconi@bs.infn.it}
\affiliation{Dipartimento di Chimica e Fisica per l'Ingegneria e per i 
Materiali, Universit\`a di Brescia, I-25123 Brescia, Italy, and\\
Istituto Nazionale di Fisica Nucleare, Sezione di Pavia, I-27100 Pavia, Italy}

\author{Marco Radici}
\email{marco.radici@pv.infn.it}
\affiliation{Dipartimento di Fisica Nucleare e Teorica, Universit\`{a} di 
Pavia, and\\
Istituto Nazionale di Fisica Nucleare, Sezione di Pavia, I-27100 Pavia, Italy}

\begin{abstract}
We perform numerical simulations of the Sivers effect from single spin asymmetries 
in Drell-Yan processes on transversely polarized protons. We consider colliding 
antiprotons and pions at different kinematic conditions of interest for the future 
planned experiments. We conventionally name "framework I" the results obtained when 
properly accounting for the various flavor dependent polarized valence 
contributions in the numerator of the asymmetry, and for the unpolarized nonvalence 
contribution in its denominator. We name "framework II" the results obtained when 
taking a suitable flavor average of the valence contributions and neglecting the 
nonvalence ones. We compare the two methods, also with respect to the input 
parametrization of the Sivers function which is extracted from data with 
approximations sometimes intermediate between frameworks I and II. Deviations between
the two approaches are found to be small except for dilepton masses below 3 GeV. The 
Sivers effect is used as a test case; the arguments can be generalized to other 
interesting azimuthal asymmetries in Drell-Yan processes, such as the Boer-Mulders
effect. 
\end{abstract}


\maketitle

\section{Introduction}
\label{sec:intro}

High-energy collisions of (polarized) hadrons represent a testground for the theory
of strong interactions, the quantum chromodynamics (QCD). In fact, several 
experiments have been performed that still await for a satisfactory interpretation of the
resulting data. Among others, in collisions of the kind $p p^{(\uparrow )} \to
h^{(\uparrow )} X$~\cite{Bunce:1976yb,Adams:1991cs,Bravar:1999rq,Adams:2003fx}, 
an azimuthally asymmetric distribution of semi-inclusively produced hadrons 
$h$ (with respect to the normal of the production plane) is observed when flipping 
the transverse spin of the target proton $p$ or of the final hadrons $h$, the 
so-called transverse single-spin asymmetry (SSA). Perturbative QCD, as it can be calculated
in the collinear massless approximation, cannot consistently accommodate these 
SSA~\cite{Kane:1978nd}, sometimes very large also at high energy. 

More recently, a series of SSA 
measurements~\cite{Airapetian:2004tw,Diefenthaler:2005gx,Avakian:2005ps,Alexakhin:2005iw} 
in semi-inclusive $l p^\uparrow \to l'\pi X$ Deep-Inelastic Scattering (SIDIS) has 
renewed the interest about the QCD spin structure of hadrons, mainly because the 
theoretical situation appears more transparent. In fact, while in hadronic collisions
like $p p^{(\uparrow )} \to h^{(\uparrow )} X$ the factorization proof is 
complicated by higher-twist correlators~\cite{Qiu:1991pp} and the power-suppressed 
asymmetry can be produced by several (overlapping) mechanisms, in SIDIS a suitable 
factorization theorem~\cite{Ji:2004wu,Collins:2004nx} allows to clearly separate 
terms with different azimuthal dependences in the leading-twist cross section. 

The main feature of this factorization proof is the possibility of going beyond the 
collinear approximation, which opens new perspectives about the explanation of the
observed SSA in terms of intrinsic transverse motion of partons inside 
hadrons, and of correlations between such intrinsic transverse momenta and 
transverse spin degrees of freedom. One of the most popular examples is the socalled 
Sivers effect~\cite{Sivers:1990cc}, where an asymmetric azimuthal distribution of 
detected hadrons (with respect to the normal to the production plane) is obtained 
from the nonperturbative correlation ${\bm p}_\sT \times {\bm P}\cdot {\bm S}_\sT$, 
with ${\bm p}_\sT$ the intrinsic transverse momentum of an unpolarized parton 
inside a target hadron with momentum ${\bm P}$ and transverse polarization 
${\bm S}_\sT$. The size of the effect is driven by a new Transverse-Momentum 
Dependent (TMD) leading-twist partonic function, the socalled Sivers function 
$f_{1\sT}^\perp$, which describes how the distribution of unpolarized partons is
distorted by the transverse polarization of the parent hadron. Then, the extraction 
of $f_{1\sT}^\perp$ would allow to study the orbital motion and the spatial
distribution of hidden confined partons, with interesting connections with the
problem of the proton spin sum rule and the powerful formalism of Generalized 
Parton Distributions~\cite{Burkardt:2003je}. 

In single-polarized Drell-Yan processes like $H_1 H_2^\uparrow \to l^+ l^- X$, there is
a situation similar to SIDIS: a suitable factorization theorem 
holds~\cite{Collins:1984kg,Collins:2004nx} and 
different asymmetric contributions can be clearly distinguished. The cross section 
must be differential in the azimuthal orientation $\phi$ of the final lepton plane 
and of the hadron polarization $\phi_S$ with respect to the reaction 
plane~\cite{Boer:1999mm}: at leading twist, it includes a term driven by 
$f_{1\sT}^\perp$ with the characteristic $\sin (\phi - \phi_S)$ dependence.
Surprisingly, there are no data for this process. At the same time, quite
interestingly the extraction of $f_{1\sT}^\perp$ from a Drell-Yan SSA would allow to
verify its predicted sign change with respect to SIDIS~\cite{Collins:2002kn}, a 
theorem based on general grounds which represents a formidable test of QCD
universality. 

Drell-Yan measurements, with unpolarized and/or transversely polarized hadrons, are 
planned by several experimental collaborations (RHIC at BNL, COMPASS at CERN, PANDA 
and PAX at GSI, and, possibly, also future experiments at JPARC). In a series of 
previous 
papers~\cite{Bianconi:2004wu,Bianconi:2005bd,Bianconi:2005yj,Bianconi:2006hc}, we 
performed numerical simulations of Drell-Yan SSA with transversely polarized protons 
using colliding protons, antiprotons, and pions, in various kinematics of interest 
for the planned experiments. In particular, we verified that the foreseen setup of
RHIC and COMPASS, with a reasonable sample of Drell-Yan events, should allow to 
unambiguously extract $f_{1\sT}^\perp$ from the corresponding $\sin (\phi -\phi_S)$ 
asymmetry, as well as to clearly test its predicted sign change with respect to the 
SIDIS asymmetry~\cite{Bianconi:2005yj,Bianconi:2006hc}. We also explored another
interesting piece of the single-polarized Drell-Yan cross 
section~\cite{Bianconi:2006hc} (see also Ref.~\cite{Sissakian:2005vd}). It is driven by 
the $\sin (\phi +\phi_S)$ asymmetry and is related to another TMD function, the 
Boer-Mulders function $h_1^\perp$, which describes the distribution of transversely 
polarized partons inside unpolarized hadrons. The interest in $h_1^\perp$ arises from
the possibility of directly linking it to the long-standing problem of the violation 
of the socalled Lam-Tung sum rule~\cite{Boer:1999mm}, namely the presence of an 
anomalous $\cos 2\phi$ asymmetry in the distribution of Drell-Yan muon pairs in
pion-induced unpolarized 
collisions~\cite{Falciano:1986wk,Guanziroli:1987rp,Conway:1989fs} (but apparently not
present in the recent data of Ref.~\cite{Zhu:2006gx} about high-energy proton-deuteron 
collisions), which neither complicated QCD calculations at higher
order, nor higher twist contributions, are able to justify in a consistent 
picture~\cite{Brandenburg:1994wf,Eskola:1994py,Berger:1979du}, and that it can
alternatively be interpreted as a QCD vacuum effect~\cite{Boer:2004mv}.

Our simulations of Drell-Yan SSA need a phenomenological input for the various TMD
functions involved. Actually, one of our goals was to test the relation between the
statistical uncertainty of simulated events and the theoretical uncertainty
originating from different parametrizations of the same TMD function. One of the
main features of these phenomenological analyses is the approximation of neglecting
both polarized and unpolarized nonvalence partons (see, e.g., 
Ref.~\cite{Anselmino:2005ea}), which actually amounts to effectively include their
contribution in the fitting parameters of the valence partons for the $x$ range
considered. In the Monte Carlo code of 
Refs.~\cite{Bianconi:2004wu,Bianconi:2005bd,Bianconi:2005yj,Bianconi:2006hc}, we
consistently take the same approach, but we further conveniently make a suitable
flavor average of the valence contribution which allows for a great simplification
of formulae. We conventionally name this scheme as "framework II". Here, we consider
also the socalled "framework I", where we release the approximation about the flavor
average and we include also the unpolarized nonvalence contribution. The goal is to 
critically discuss the two methods, also 
with respect to the approximated framework introduced by those parametrization of the
Sivers function that are somewhat intermediate between them. More specifically,
we want to identify and quantify the deviations of the results obtained within  
"framework I" from those obtained within 
"framework II". At a qualitative level, the origin of these deviations can be easily
identified {\it a priori}, while from the quantitative point of view there are
quite distinct situations that need to be separately analyzed. We will systematically
adopt the Sivers effect as our test case, because the relative abundance of these 
SSA data allows to already build realistic parametrizations of $f_{1\sT}^\perp$, 
some of which are discussed in Sec.~\ref{sec:formulae}. However, the argument 
can be generalized, in principle, also to the azimuthal asymmetry generated by 
the Boer-Mulders effect or by the violation of the Lam-Tung sum rule.

The paper is organized as follows. In Sec.~\ref{sec:formulae}, the general formalism,
the considered phenomenological parametrizations of the Sivers function, and the main
approximations leading to the definition of "framework I" and "framework II", are 
discussed. In Sec.~\ref{sec:mcout}, simulations for SSA within the two frameworks 
are compared for different Drell-Yan collisions on transversely polarized proton 
targets in several kinematic conditions of interest.
Finally, in Sec.~\ref{sec:end} some conclusions are drawn.

\begin{figure}[h]
\centering
\includegraphics[width=7cm]{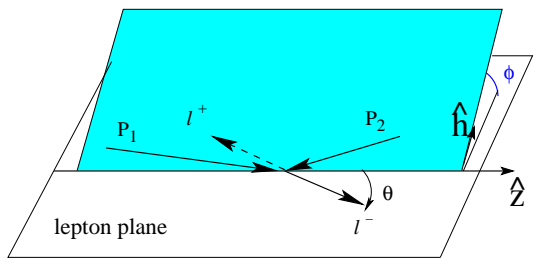}
\caption{The Collins-Soper frame.}
\label{fig:dyframe}
\end{figure}

\section{General formalism and approximations}
\label{sec:formulae}

In a Drell-Yan process, an antilepton and a lepton with individual momenta $k_1$ 
and $k_2$ are produced from the collision of two hadrons with momentum $P_i$, mass 
$M_i$, and spin $S_i$, with $i=1,2$. The center-of-mass (c.m.) square energy 
available is $s=(P_1+P_2)^2$ and the invariant mass of the final pair is 
given by the time-like momentum transfer $q^2 \equiv M^2 = (k_1 + k_2)^2$. If 
$M^2,s \rightarrow \infty$, while keeping the ratio $0\leq \tau = M^2/s \leq 1$ 
limited, the factorized elementary mechanism proceeds through the annihilation of a 
parton and an antiparton with momenta $p_1$ and $p_2$, respectively, into a virtual 
photon with time-like momentum $q^2$. If $P_1^+$ and $P_2^-$ are the dominant 
light-cone components of hadron momenta in this regime, then the partons are 
approximately collinear with the parent hadrons and carry the light-cone momentum 
fractions $0\leq x_1 = p_1^+ / P_1^+ , \; x_2 = p_2^- / P_2^- \leq 1$, with 
$q^+ = p_1^+, \; q^- = p_2^-$ by momentum conservation~\cite{Boer:1999mm}. The 
transverse components ${\bm p}_{i\sT}$ of $p_i$ with respect to the direction 
defined by ${\bm P}_i (i=1,2)$, are constrained again by the momentum conservation 
${\bm q}_\sT = {\bm p}_{1\sT} + {\bm p}_{2\sT}$, where ${\bm q}_\sT$ is the 
transverse momentum of the final lepton pair. If ${\bm q}_\sT \neq 0$ the 
annihilation direction is not known. Hence, it is convenient to select the 
socalled Collins-Soper frame~\cite{Collins:1977iv} described in 
Fig.~\ref{fig:dyframe}. The final lepton pair is detected in the solid angle 
$(\theta, \phi )$, where, in particular, $\phi$ (and all other azimuthal angles) 
is measured in a plane perpendicular to the indicated lepton plane but containing 
$\hat{\bm h} = {\bm q}_\sT / |{\bm q}_\sT|$. 

By neglecting terms $\sim 1/M^2$, with $M$ the largest mass of the initial hadrons, 
the expression of the leading-twist differential cross section for the $H_1
H_2^\uparrow \to l^+ l^- X$ process can be written as~\cite{Boer:1999mm}
\bea
\frac{d\sigma}{d\Omega dx_1 dx_2 d{\bm q}_\sT} &= &
\frac{d\sigma^o}{d\Omega dx_1 dx_2 d{\bm q}_\sT} + 
\frac{d\Delta \sigma^\uparrow}{d\Omega dx_1 dx_2 d{\bm q}_\sT}  \nn \\
&= &\frac{\alpha^2}{3Q^2}\,\sum_q\,e_q^2\,\Bigg\{ A(y) \, 
{\cal F}\left[ f_1^q(H_1)\, f_1^q (H_2) \right] \nn \\
& &\mbox{\hspace{2cm}} + B(y) \, \cos 2\phi \, 
{\cal F}\left[ \left( 2 \hat{\bm h}\cdot {\bm p}_{1\sT} \, \hat{\bm h} \cdot 
{\bm p}_{2\sT} - {\bm p}_{1\sT} \cdot {\bm p}_{2\sT} \right) \, 
\frac{h_1^{\perp\,q}(H_1)\,h_1^{\perp\,q}(H_2)}{M_1\,M_2}\,\right] \Bigg\} \nn \\
& &+ \frac{\alpha^2}{3Q^2}\,|{\bm S}_{2\sT}|\,\sum_q\,e_q^2\,\Bigg\{ 
A(y) \, \sin (\phi - \phi_{S_2})\, {\cal F}\left[ \hat{\bm h}\cdot 
{\bm p}_{2\sT} \,\frac{f_1^q(H_1) \, f_{1\sT}^{\perp\,q}(H_2^\uparrow)}{M_2}\right] 
\nn \\
& &\mbox{\hspace{3cm}} - B(y) \, \sin 
(\phi + \phi_{S_2})\, {\cal F}\left[ \hat{\bm h}\cdot {\bm p}_{1\sT} \,
\frac{h_1^{\perp\,q}(H_1) \, h_1^q(H_2^\uparrow)}{M_1}\right] \, \Bigg\} \; ,
\label{eq:xsect}
\eea
where $\alpha$ is the fine structure constant, $d\Omega = \sin \theta d\theta
d\phi$, $e_q$ is the charge of the parton with flavor $q$, $\phi_{S_2}$ is
the azimuthal angle of the transverse polarization vector of the hadron 
$H_2^\uparrow$, and 
\begin{align}
A(y) = \left( \frac{1}{2} - y + y^2 \right) \, \stackrel{\mbox{cm}}{=}\, 
\frac{1}{4}\left( 1 + \cos^2 \theta \right) &\mbox{\hspace{2cm}} 
B(y) = y (1-y) \, \stackrel{\mbox{cm}}{=}\,\frac{1}{4}\, \sin^2 \theta \; . 
\label{eq:lepton}
\end{align}
The TMD functions $f_1^q(H), \, h_1^{\perp\,q}(H)$, describe the distributions 
of unpolarized and transversely polarized partons in an unpolarized hadron $H$,
respectively, while $f_{1\sT}^{\perp\,q}(H^\uparrow)$ and $h_1^q (H^\uparrow)$ 
have a similar interpretation but for transversely polarized hadrons $H^\uparrow$. 
The convolutions are defined as 
\bea
{\cal F} \left[ \mbox{TMD}_1^q(H_1) \, \mbox{TMD}_2^q(H_2^{(\uparrow )}) \right] 
&\equiv &\int d{\bm p}_{1\sT} d{\bm p}_{2\sT}\, \delta \left( {\bm p}_{1\sT} + 
{\bm p}_{2\sT} - {\bm q}_\sT \right) \nn \\
& &\mbox{\hspace{2cm}} \times \left[ \mbox{TMD}_1(x_1,{\bm p}_{1\sT}; 
\bar{q}/H_1)\, \mbox{TMD}_2(x_2,{\bm p}_{2\sT}; q/H_2^{(\uparrow )} ) + 
(q\leftrightarrow \bar{q}) \right] \; .
\label{eq:convol}
\eea

The Monte Carlo events have been generated by the following cross 
section~\cite{Bianconi:2004wu}:
\be
\frac{d\sigma}{d\Omega dx_1 dx_2 d{\bm q}_\sT} = K \, \frac{1}{s}\, 
|{\cal T}({\bm q}_\sT, x_1, x_2, M)|^2 \, \sum_{i=1}^4\, c_i ({\bm q}_\sT, 
x_1,x_2) \, S_i(\theta, \phi, \phi_{_{S_2}}) \; ,
\label{eq:mc-xsect}
\ee
where the event distribution is driven by the elementary unpolarized annihilation,
whose transition amplitude ${\cal T}$ has been highlighted. In
Eq.~(\ref{eq:xsect}), we assume a factorized transverse-momentum dependence in 
each TMD such as to break the convolution ${\cal F}$, leading to
\be
|{\cal T}|^2 \approx A(q_\sT,x_1,x_2,M) \, F(x_1,x_2) \; ,
\label{eq:factorized}
\ee
where $q_\sT \equiv |{\bm q}_\sT|$. The function $A$ is parametrized and
normalized as in Ref.~\cite{Conway:1989fs}, where high-energy Drell-Yan $\pi - p$ 
collisions were considered. The average transverse momentum turns out to be 
$\langle q_\sT \rangle > 1$ GeV/$c$ (see also the more recent 
Ref.~\cite{Towell:2001nh}), which effectively reproduces the influence of sizable
QCD corrections beyond the parton model picture of Eq.~(\ref{eq:xsect}). It is
well known~\cite{Altarelli:1979ub} that such corrections induce also large $K$
factors and an $M$ scale dependence in parton distributions, determining their
evolution. As in our previous 
works~\cite{Bianconi:2004wu,Bianconi:2005yj,Bianconi:2005bd,Bianconi:2006hc}, we 
conventionally assume in Eq.~(\ref{eq:mc-xsect}) that $K=2.5$, but we stress that 
in a single-spin asymmetry the corrections to the cross sections in the numerator 
and in the denominator should compensate each other, as it turns out to actually 
happen at RHIC c.m. square energies~\cite{Martin:1998rz}. Since the range of $M$ 
values here explored is close to the one of Ref.~\cite{Conway:1989fs}, where the 
parametrization of $A, F,$ and $c_i$ in Eq.~(\ref{eq:mc-xsect}), was deduced 
assuming $M$-independent parton distributions, we keep our same previous
approach~\cite{Bianconi:2004wu,Bianconi:2005yj,Bianconi:2005bd,Bianconi:2006hc} 
and use
\be
F(x_1,x_2) = \frac{\alpha^2}{12 Q^2}\,\sum_q\,e_q^2\,
f_1^q(x_1; \bar{q}/H_1) \, f_1^q (x_2; q/H_2) + (\bar{q} \leftrightarrow q) \; , 
\label{eq:mcF}
\ee
where the unpolarized distribution $f_1^q (x)$ for various flavors $q=u,d,s,$ is 
taken again from Ref.~\cite{Conway:1989fs}.

The whole solid angle $(\theta, \phi)$ of the final lepton pair in the 
Collins-Soper frame is randomly distributed in each variable. The explicit form 
for sorting it in the Monte-Carlo
is~\cite{Bianconi:2004wu,Bianconi:2005yj,Bianconi:2006hc}
\bea
\sum_{i=1}^4\, c_i (q_\sT,x_1,x_2) \, S_i(\theta, \phi, \phi_{S_2}) 
&= &(1 + \cos^2 \theta) \nn \\
& &+ \frac{\nu (q_\sT,x_1,x_2)}{2}\, \sin^2\theta \, \cos 2\phi \nn \\
& &+ |{\bm S}_{2\sT}|\, c_{Siv} (q_\sT,x_1,x_2)\, (1+\cos^2 \theta) \, 
\sin (\phi - \phi_{S_2}) \nn \\
& &+ |{\bm S}_{2\sT}|\, c_{BM} (q_\sT,x_1,x_2)\, \sin^2 \theta \, 
\sin (\phi + \phi_{S_2}) \; . 
\label{eq:mcS}
\eea
If quarks were massless, the virtual photon would be only transversely polarized 
and the angular dependence would be described by the functions $c_1 = 1$ 
and $S_1 = 1 + \cos^2 \theta$. Violations of such azimuthal symmetry 
induced by the function $c_2 \equiv \textstyle{\frac{\nu}{2}}$ are due to the 
longitudinal polarization of the virtual photon and to the fact that quarks have 
an intrinsic transverse momentum distribution, leading to the explicit dependence of
$\nu$ upon $q_\sT$ and to the violation of the socalled Lam-Tung sum 
rule~\cite{Falciano:1986wk,Guanziroli:1987rp,Conway:1989fs}. QCD corrections 
influence $\nu$, which in principle depends also on $M^2$~\cite{Conway:1989fs}. 
Azimuthal $\cos 2\phi$ asymmetries were simulated in Ref.~\cite{Bianconi:2004wu} using 
the simple parametrization of Ref.~\cite{Boer:1999mm} and testing it against the 
previous measurement of Ref.~\cite{Falciano:1986wk,Guanziroli:1987rp,Conway:1989fs}. 

The next term in Eq.~(\ref{eq:mcS}) describes the Sivers 
effect~\cite{Sivers:1990cc}:
\be
c_3 \equiv c_{Siv} (q_\sT,x_1,x_2) = 
\frac{\sum_q\,e_q^2\,{\cal F}\left[ \hat{\bm h}\cdot {\bm p}_{2\sT} \, 
\displaystyle{\frac{f_1^q(x_1,{\bm p}_{1\sT})\, 
f_{1T}^{\perp\, q}(x_2,{\bm p}_{2\sT})}{M_2}} \right]}
{\sum_q\,e_q^2\,{\cal F}\left[ f_1^q(x_1,{\bm p}_{1\sT})\, 
 f_1^q(x_2,{\bm p}_{2\sT}) \right]} \; ,
\label{eq:mcc-sivers}
\ee
while the last one contains the asymmetry induced by the socalled Boer-Mulders
effect~\cite{Boer:1999mm}:
\be
c_4 \equiv c_{BM} (q_\sT,x_1,x_2) = - 
\frac{\sum_q\,e_q^2\,{\cal F}\left[ \hat{\bm h}\cdot {\bm p}_{1\sT} \, 
\displaystyle{\frac{h_1^{\perp\, q}(x_1,{\bm p}_{1\sT})\, 
h_1^q(x_2,{\bm p}_{2\sT})}{M_1}} \right]}
{\sum_q\,e_q^2\,{\cal F}\left[ f_1^q(x_1,{\bm p}_{1\sT})\, 
 f_1^q(x_2,{\bm p}_{2\sT}) \right]} \; .
\label{eq:mcc-boer}
\ee

In the following, we will consider the asymmetry generated only by $c_{Siv}$ in
Eq.~(\ref{eq:mcc-sivers}), because there is a sufficient amount of available data 
to support the construction of realistic parametrizations for the Sivers function 
$f_{1T}^{\perp}$. However, our arguments can be easily generalized also to the 
Boer-Mulders $c_{BM}$ term. We will come back on the $\nu$ coefficient at the end 
of next Section.

For sake of consistency, the denominator of Eq.~(\ref{eq:mcc-sivers}) is
approximated by the same $|{\cal T}|^2$ of Eq.~(\ref{eq:factorized}). As for the
numerator, we first simulate the Sivers effect using the parametrization of 
Ref.~\cite{Anselmino:2005ea}, 
\bea
f_{1T}^{\perp\, q}(x,{\bm p}_\sT) &= &-2\, N_q\,
\frac{(a_q+b_q)^{a_q+b_q}}{a_q^{a_q}\,b_q^{b_q}}\,
x^{a_q}\,(1-x)^{b_q}\,\frac{M_2 M_0}{{\bm p}_\sT^2+M_0^2}\,
f_1^q(x,{\bm p}_\sT) \nn \\
&= &-2\, N_q\,\frac{1}{\pi \, \langle p_\sT^2 \rangle}\,
\frac{(a_q+b_q)^{a_q+b_q}}{a_q^{a_q}\, b_q^{b_q}} \, x^{a_q}\, (1-x)^{b_q}\, 
\frac{M_2 M_0}{{\bm p}_\sT^2+M_0^2}\, e^{-p_\sT^2/\langle p_\sT^2 \rangle}\, 
f_1^q(x) \; ,
\label{eq:pTanselm}
\eea
where $M_2$ is the mass of the polarized proton, $p_\sT \equiv |{\bm p}_\sT|$, 
and $\langle p_\sT^2 \rangle = 0.25$ (GeV/$c$)$^2$ is deduced by assuming a 
Gaussian ansatz for the ${\bm p}_\sT$ dependence of $f_1$ in order to reproduce the 
azimuthal angular dependence of the SIDIS unpolarized cross section (Cahn effect). 
Flavor-dependent normalization and parameters in the $x$ dependence are fitted to 
SIDIS SSA data using only the two flavors $q=u,d$ and neglecting the (small) 
contribution of antiquarks (see Refs.~\cite{Anselmino:2005ea,Bianconi:2006hc}).

Following Ref.~\cite{Boer:1999mm} and including a sign change of $f_{1T}^{\perp}$
when plugging it in the Drell-Yan cross section~\cite{Collins:2002kn}, we get
\bea
c_{Siv}^A &\approx &\frac{4 M_0\,q_\sT}{q_\sT^2+4 M_0^2}\, 
\frac{\sum_{q}\,e_q^2\,N_q\,
  \displaystyle{ \frac{(\alpha_q+\beta_q)^{\alpha_q+\beta_q}}
                   {\alpha_q^{\alpha_q}\,\beta_q^{\beta_q}} }\,
\bar{f}_1^q(x_1)\, x_2^{\alpha_q}\,(1-x_2)^{\beta_q}\, f_1^q(x_2)}
{\sum_q\,e_q^2\,\bar{f}_1^q(x_1)\,f_1^q(x_2)+ (1\leftrightarrow 2)} \nn \\
&\equiv &N_A(q_\sT)\, 
\frac{e_u^2 \, \bar{u}(x_1)\,u_{Siv}^A(x_2) + e_d^2\,\bar{d}(x_1)\,d_{Siv}^A(x_2)}
{\left[ e_u^2\,\bar{u}(x_1)\,u(x_2) + e_d^2\,\bar{d}(x_1)\,d(x_2) + 
e_s^2\,\bar{s}(x_1)\,s(x_2)\right] + (1 \leftrightarrow 2)} \; ,
\label{eq:csiv-A}
\eea
where, for brevity, $u_{Siv}^A(x)$ represents the contribution of flavor $u$ to 
the $x$ dependence of the Sivers function parametrized as in 
Eq.~(\ref{eq:pTanselm}) (and similarly for flavor $d$). 

As an alternative choice, we adopt the new parametrization described in
Ref.~\cite{Bianconi:2005yj,Bianconi:2006hc}. It is inspired to the one of 
Ref.~\cite{Vogelsang:2005cs}, whose $x$ dependence is retained but a different 
flavor-dependent normalization and an explicit ${\bm p}_\sT$ dependence are 
introduced. The latter is bound to the shape of the recent RHIC data on 
$pp^\uparrow \to \pi X$ at $\sqrt{s}=200$ GeV~\cite{Adler:2005in}, where
large persisting asymmetries are found that could be partly due to the 
leading-twist Sivers mechanism. The expression adopted is
\bea
f_{1T}^{\perp\, q}(x,{\bm p}_\sT) &=&N_q\,x\,(1-x)\,
\frac{M_2 p_0^2 p_\sT}{(p_\sT^2+\frac{p_0^2}{4})^2}\,f_1^q(x,{\bm p}_\sT) \nn \\
&= & N_q\,x\,(1-x)\,\frac{M_2 p_0^2 p_\sT}{(p_\sT^2+\frac{p_0^2}{4})^2}\,
\frac{1}{\pi \, \langle p_\sT^2 \rangle}\, e^{-p_\sT^2/\langle p_\sT^2 \rangle}\, 
f_1^q(x) \; ,
\label{eq:pTnoi}
\eea
where $p_0 = 2$ GeV/$c$, and $N_u = - N_d = 0.7$. The sign, positive for $u$
quarks and negative for the $d$ ones, already takes into account the predicted sign
change of $f_{1\sT}^\perp$ from Drell-Yan to SIDIS~\cite{Collins:2002kn}. 

Along the same previous lines, we get
\bea
c_{Siv}^B &\approx &\left( \frac{2\, p_0 \, q_\sT}{q_\sT^2+p_0^2} \right)^2\, 
\frac{\sum_{q}\,e_q^2\,N_q\,\bar{f}_1^q(x_1)\, x_2\,(1-x_2)\, f_1^q(x_2)}
{\sum_q\,e_q^2\,\bar{f}_1^q(x_1)\,f_1^q(x_2)+ (1\leftrightarrow 2)} \nn \\
&\equiv &N_B(q_\sT)\, 
\frac{e_u^2 \, \bar{u}(x_1)\,u_{Siv}^B(x_2) + e_d^2\,\bar{d}(x_1)\,d_{Siv}^B(x_2)}
{\left[ e_u^2\,\bar{u}(x_1)\,u(x_2) + e_d^2\,\bar{d}(x_1)\,d(x_2) + 
e_s^2\,\bar{s}(x_1)\,s(x_2)\right] + (1 \leftrightarrow 2)} \; ,
\label{eq:csiv-B}
\eea
where now $u_{Siv}^B(x)$ indicates the contribution of flavor $u$ to 
the $x$ dependence of the Sivers function parametrized as in 
Eq.~(\ref{eq:pTnoi}).

In the following, we will refer to "framework I" as to the 
$(1+\cos^2 \theta)\,\sin (\phi-\phi_{S_2})$ 
angular asymmetry generated in Eq.~(\ref{eq:xsect}) by the 
coefficients~(\ref{eq:csiv-A}) or~(\ref{eq:csiv-B}).  

The coefficients $c_{Siv}^A$ and $c_{Siv}^B$ can be further approximated with a
procedure that here we will conventionally indicate as "framework II". Again, 
following the lines described in Refs.~\cite{Bianconi:2005yj,Bianconi:2006hc} we 
obtain
\bea
c_{Siv}^A &= &N_A(q_\sT)\, 
\frac{e_u^2 \, \bar{u}(x_1)\,u_{Siv}^A(x_2) + e_d^2\,\bar{d}(x_1)\,d_{Siv}^A(x_2)}
{\left[ e_u^2\,\bar{u}(x_1)\,u(x_2) + e_d^2\,\bar{d}(x_1)\,d(x_2) + 
e_s^2\,\bar{s}(x_1)\,s(x_2)\right] + (1 \leftrightarrow 2)} \nn \\
&\approx &N_A(q_\sT)\, \left[ n_u^A\,\frac{u_{Siv}^A(x_2)}{u(x_2)} + 
n_d^A\,\frac{d_{Siv}^A(x_2)}{d(x_2)} \right] \; ,
\label{eq:csiv-Abis}
\eea
and
\bea
c_{Siv}^B &= &N_B(q_\sT)\, 
\frac{e_u^2 \, \bar{u}(x_1)\,u_{Siv}^B(x_2) + e_d^2\,\bar{d}(x_1)\,d_{Siv}^B(x_2)}
{\left[ e_u^2\,\bar{u}(x_1)\,u(x_2) + e_d^2\,\bar{d}(x_1)\,d(x_2) + 
e_s^2\,\bar{s}(x_1)\,s(x_2)\right] + (1 \leftrightarrow 2)} \nn \\
&\approx &N_B(q_\sT)\, \left[ n_u^B\,\frac{u_{Siv}^B(x_2)}{u(x_2)} + 
n_d^B\,\frac{d_{Siv}^B(x_2)}{d(x_2)} \right] \; .
\label{eq:csiv-Bbis}
\eea
The coefficients $n_q$ ($q=u,d$) include the contribution of the quark charge, of
the normalization of the parton distributions, as well as of the statistical weight
of the considered flavor. In fact, the approximation is based on the idea that each
term in both flavor sums in the numerator and denominator can be replaced by a 
"flavor-averaged" one, and the resulting simplified ratio is then properly 
weighted. For example, in a $\bar{p}p^\uparrow$ collision where only
valence contributions are considered and the $u(x)$ distribution in $p$ is 
normalized to 2 (and similarly for the $\bar{u}$ one in $\bar{p}$), we easily get a
statistical ratio 16:1 of the $\bar{u}u^\uparrow$ annihilations over the 
$\bar{d}d^\uparrow$ ones, such that $n_u=16/17,\, n_d = 1/17$.


The underlying idea in "framework II" is that there is no strong flavor dependence in
the sums appearing in Eqs.~(\ref{eq:csiv-A}) and (\ref{eq:csiv-B}), and that it is
possible to neglect the contribution from sea (anti)quarks. The latter feature is 
included by construction in the parametrizations of the Sivers functions, in
agreement with the common belief about the behaviour of SSA at low $x$. It is
far less obvious that the approximation can be safely carried on also in the
denominator of the asymmetries $c_{Siv}^{A/B}$, where the unpolarized distributions
$f_1$ are involved. In the next Section, we will numerically simulate 
$\pi^\pm p^\uparrow $ and $\bar{p}p^\uparrow$ Drell-Yan collisions to explore the 
effect of neglecting such terms.

\section{Monte Carlo simulation and discussion of results}
\label{sec:mcout}

In this Section, we present results for the Monte Carlo simulation of the Sivers effect
in several Drell-Yan events using transversely polarized proton targets by adopting
either "framework I" or "framework II", as they are described in the previous Section. The 
goal is to explore the sensitivity of the results when the contribution of unpolarized 
sea (anti)quarks is neglected in the denominator of the SSA, as it is 
usually done when parametrizing the Sivers function from experimental data. 

In the Monte Carlo, events are simulated by the cross section~(\ref{eq:mc-xsect}) with 
"framework II", namely using Eqs.~(\ref{eq:csiv-Abis}) or (\ref{eq:csiv-Bbis}) according 
to the input parametrization selected for the Sivers 
function~\cite{Anselmino:2005ea,Bianconi:2005yj,Bianconi:2006hc}; the unpolarized 
parton distributions are parametrized following Ref.~\cite{Conway:1989fs}, as explained 
in the previous Section when discussing Eqs.~(\ref{eq:factorized}) and (\ref{eq:mcF}). 
Events for the Sivers effect are then rejected/accepted by constructing the spin
asymmetry $A'=(U'-D')/(U'+D')$, where $U'$ ($D'$) represents events with positive (negative) 
values of $\sin (\phi - \phi_{S_2})$ in Eq.~(\ref{eq:mcS}). Similarly, the spin 
asymmetry $A=(U-D)/(U+D)$ is obtained by rejecting/accepting events using the 
"framework I", namely using Eqs.~(\ref{eq:csiv-A}) or (\ref{eq:csiv-B}), for positive 
($U$) or negative ($D$) values of $\sin (\phi - \phi_{S_2})$. Data are accumulated 
only in the $x_2$ bins of the polarized proton, i.e. they are summed over in the 
$x_1$ bins for the hadronic beam, in the transverse momentum $q_\sT$ of the
lepton pair 
and in their zenithal orientation $\theta$. In the following, plots will compare 
$A$ and $A'$ for different Drell-Yan processes and kinematical conditions,
showing only the positive values of the asymmetries (the negative ones look 
equally distributed in a symmetric way).

All events refer to collisions at $s=100$ GeV$^2$, which can be explored either at GSI
in the socalled collider mode, or at COMPASS with fixed targets. The lepton pair 
invariant mass is constrained in the ranges $4<M<6$ GeV and $1.5<M<2.5$ GeV in order to 
avoid overlaps with the resonances of the $\bar{c}c$ and $\bar{b}b$ quarkonium 
systems, while exploring at the same time different $\langle x_2 \rangle$ regions which
stress the role of sea (anti)quarks. The transverse momentum of the lepton pair is 
constrained in the range $1<q_\sT <3$ GeV/$c$ in order to avoid a strong dilution of 
the SSA because of the rapid decrease of the distributions~(\ref{eq:pTanselm}) and
(\ref{eq:pTnoi}) at larger $q_\sT$. Moreover, the resulting $\langle q_\sT \rangle 
\sim 1.8$ GeV/$c$ is in fair agreement with the one experimentally explored at 
RHIC~\cite{Adler:2005in}. 

\begin{figure}[h]
\centering
\includegraphics[width=7cm]{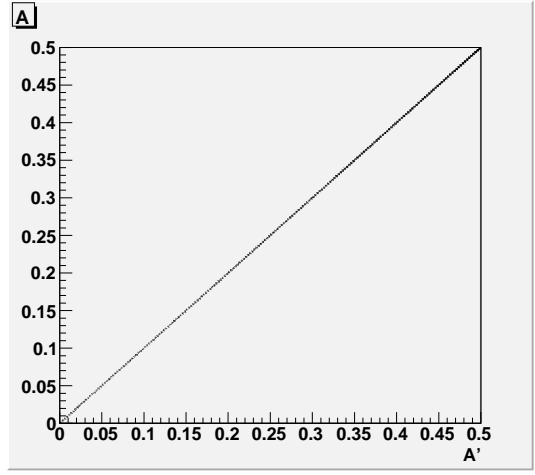}
\caption{Scatter plot for single-spin asymmetries $A$ and $A'$, calculated with
"framework I" and "framework II" (see text), produced by $7\, 000$ events of the 
Sivers effect in the $\bar{p} p^\uparrow \to \mu^+ \mu^- X$ process with muon 
invariant mass $1.5<M<2.5$ GeV and transverse momentum $1<q_\sT <3$ GeV/$c$, where 
only (polarized) valence contributions are considered and 
$u_{Siv}(x) \equiv u(x) , \, d_{Siv}(x) \equiv d(x)$.}
\label{fig:antip-siv1}
\end{figure}

The considered Drell-Yan collisions involve transversely polarized protons and 
different hadronic probes: antiprotons ($\bar{p}$) and pions ($\pi^-$ and $\pi^+$). The
statistical sample is made of $7\, 000$ events except for the $\pi^+$ probe, where we
have used $20\, 000$ events because the Monte Carlo indicates that the cross section 
involving $\pi^+$ is statistically disfavoured by approximately a factor 
1/4~\cite{Bianconi:2005bv}.

As a sort of "warm up", we first consider oversimplified cases where the simulated
comparison between $A$ and $A'$ can be confronted with {\it a priori} known results. In
Fig.~\ref{fig:antip-siv1}, we show the scatter plot of $A$ versus $A'$ for the 
$\bar{p} p^\uparrow \to \mu^+ \mu^- X$ process with $1.5<M<2.5$ GeV where only
(polarized) valence contributions are considered and 
$u_{Siv}(x) \equiv u(x) , \, d_{Siv}(x) \equiv d(x)$. From Eq.~(\ref{eq:csiv-Abis}), it
is evident that both "framework I" and "framework II" must give the same result, namely
\bea
c_{Siv}^I &\propto &
\frac{e_u^2 \, \bar{u}(x_1)\,u_{Siv}(x_2) + e_d^2\,\bar{d}(x_1)\,d_{Siv}(x_2)}
{e_u^2\,\bar{u}(x_1)\,u(x_2) + e_d^2\,\bar{d}(x_1)\,d(x_2)} \equiv 1 \nn \\
c_{Siv}^{II} &\propto &\left[ n_u\,\frac{u_{Siv}(x_2)}{u(x_2)} + 
n_d\,\frac{d_{Siv}(x_2)}{d(x_2)} \right] \equiv n_u + n_d = 1\; ,
\label{eq:csiv-test1}
\eea
because any flavor dependence has been switched off and for "framework I" the asymmetry 
$A$ is not diluted by the contribution of unpolarized sea (anti)quarks showing up in 
the denominator of the first equality in Eq.~(\ref{eq:csiv-Abis}) itself (the socalled 
"sea dilution effect"). From the figure we get the consistent picture that the SSA  
calculated with the two methods are statistically equal. We have checked 
that the same result holds also for $\pi^\pm$ probes, as it is obvious.

\begin{figure}[h]
\centering
\includegraphics[width=7cm]{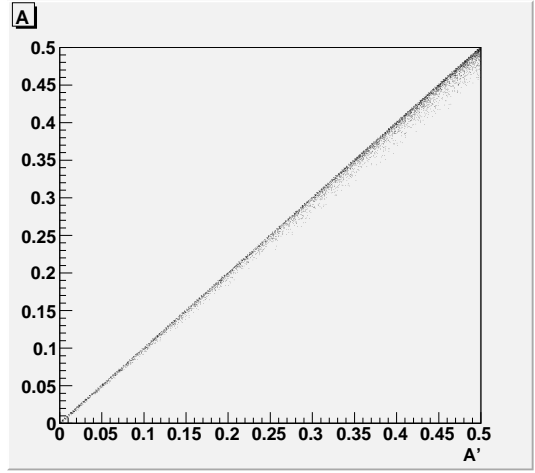}
\caption{Scatter plot in the same conditions as previous figure, but for the 
$\pi^- p^\uparrow \to \mu^+ \mu^- X$ process with muon invariant mass $4<M<6$ GeV 
and including the contribution of unpolarized sea (anti)quarks.}
\label{fig:pi-sea}
\end{figure}

In Fig.~\ref{fig:pi-sea}, the scatter plot is shown for the 
$\pi^- p^\uparrow \to \mu^+ \mu^- X$ process with $4<M<6$ GeV where still 
$u_{Siv}(x) \equiv u(x) , \, d_{Siv}(x) \equiv d(x)$, but the contribution of the
unpolarized sea (anti)quarks is included. Therefore, we have
\bea
c_{Siv}^I &\propto &
\frac{e_u^2 \, \bar{u}(x_1)\,u_{Siv}(x_2) + e_d^2\,\bar{d}(x_1)\,d_{Siv}(x_2)}
{\left[ e_u^2\,\bar{u}(x_1)\,u(x_2) + e_d^2\,\bar{d}(x_1)\,d(x_2) + 
e_s^2\,\bar{s}(x_1)\,s(x_2)\right] + (1 \leftrightarrow 2)} \nn \\
&\equiv &\frac{e_u^2 \, \bar{u}(x_1)\,u(x_2) + e_d^2\,\bar{d}(x_1)\,d(x_2)}
{\left[ e_u^2\,\bar{u}(x_1)\,u(x_2) + e_d^2\,\bar{d}(x_1)\,d(x_2) + 
e_s^2\,\bar{s}(x_1)\,s(x_2)\right] + (1 \leftrightarrow 2)} \nn \\
&= &\left[ 1 + 
\frac{e_u^2\,u(x_1)\,\bar{u}(x_2) + e_d^2\,d(x_1)\,\bar{d}(x_2) + 
e_s^2\,\bar{s}(x_1)\,s(x_2) + e_s^2\,s(x_1)\,\bar{s}(x_2)}
{e_u^2\,\bar{u}(x_1)\,u(x_2) + e_d^2\,\bar{d}(x_1)\,d(x_2)} \right]^{-1} < 1 
\label{eq:csivI-test2} \\
c_{Siv}^{II} &\propto &\left[ n_u\,\frac{u_{Siv}(x_2)}{u(x_2)} + 
n_d\,\frac{d_{Siv}(x_2)}{d(x_2)} \right] \equiv n_u + n_d = 1\; .
\label{eq:csivII-test2}
\eea
Evidently, the difference between the two approaches stems from the unpolarized
sea-(anti)quark contribution, both in the numerator and in the denominator of 
$c_{Siv}^I$, and it is responsible for the (limited) spreading of some events in the 
scatter plot.

\begin{figure}[h]
\centering
\includegraphics[width=7cm]{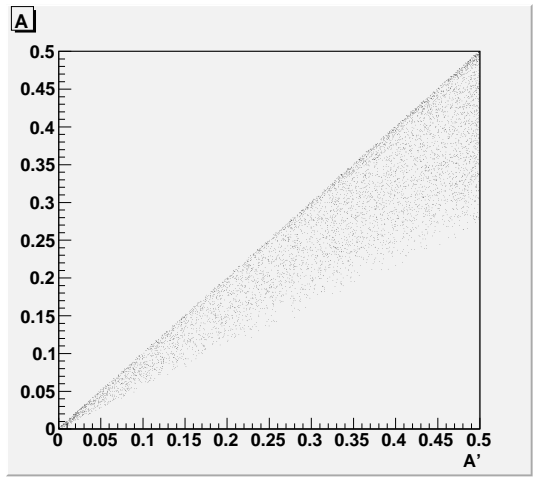}
\caption{Scatter plot in the same conditions as previous figure, but for muon 
invariant mass $1.5<M<2.5$ GeV.}
\label{fig:pi-sealowM}
\end{figure}

In Fig.~\ref{fig:pi-sealowM}, the same situation is shown for muon invariant masses
$1.5<M<2.5$ GeV. At the given $s=100$ GeV$^2$, lower $M$ implies testing lower $x_2$
values, typically below 0.05-0.1, where sea effects start dominating over the valence 
contribution. Consequently, the "sea dilution effect" becomes stronger and 
the marked difference between "framework I" and "framework II" produces the scattering of
events observed in the plot. In summary, the two approaches become equivalent when
the polarized distributions weakly depend on flavor (or, equivalently, one flavor 
dominates the flavor sum in the numerator of the SSA), and the
contribution of sea (anti)quarks becomes negligible, i.e. for not too low 
$\langle x \rangle$ and, consequently, $M$ (assuming that also transversely polarized
sea (anti)quarks can be neglected, as well).

\begin{figure}[h]
\centering
\includegraphics[width=7cm]{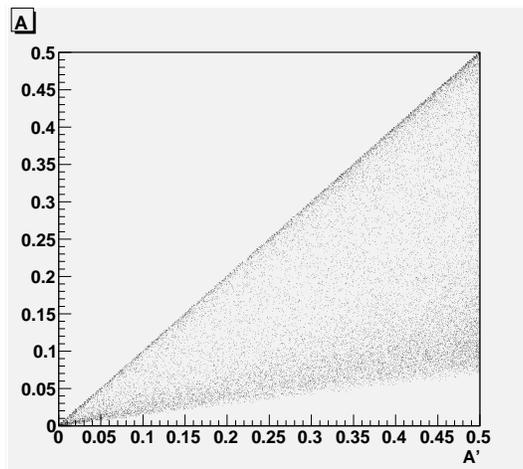}
\caption{Scatter plot in the same conditions as previous figure, but for the 
$\pi^+ p^\uparrow \to \mu^+ \mu^- X$ process.}
\label{fig:pi+sealowM}
\end{figure}

In Fig.~\ref{fig:pi+sealowM}, the same kinematical situation of the previous figure is
considered for $\pi^+$ probes. Equations~(\ref{eq:csivI-test2}) and
(\ref{eq:csivII-test2}) formally still hold. However, while the latter remains 
obviously unchanged, the former can be usefully compared in the two cases by adopting
some simple approximations. In fact, if we neglect the product of two sea-quark
distributions and we assume isospin symmetry among the valence distributions of the
$\pi^\pm$ probes (i.e., $u_{\pi^+}(x) = \bar{u}_{\pi^-}(x) = u_\pi(x) \equiv u(\pi)$ 
and $\bar{d}_{\pi^+}(x) = d_{\pi^-}(x) = d_\pi(x) \equiv d(\pi)$), we get for the 
$\pi^-$ probe
\be
c_{Siv\, \pi^-}^I \propto \left[ 1 + 
\frac{d(\pi)\,\bar{d}(p)}{u(\pi)\,u(p) + \bar{d}(\pi^-)\,d(p)} \right]^{-1}
\label{eq:csivI-test2-b} \; ,
\ee
and for the $\pi^+$ probe
\be
c_{Siv\, \pi^+}^I \propto \left[ 1 + 
\frac{4 u(\pi)\,\bar{u}(p)}{d(\pi)\,d(p) + 8 \bar{u}(\pi^+)\,u(p)} \right]^{-1} \; .
\label{eq:csivI-test2-c}
\ee
For sake of simplicity, we further assume the flavor independence of the sea 
contributions, namely 
$\bar{d}(\pi^-) = \bar{u}(\pi^+) \equiv q_{sea}(\pi) \ll u(\pi), \, d(\pi)$ and
$\bar{u}(p) = \bar{d}(p) \equiv q_{sea}(p) \ll u(p), \, d(p)$. It is also reasonable
to consider $u(\pi) > d(\pi)$ and $u(p) > d(p)$. From these inequalities it follows
that 
\be
\frac{d(\pi)\,q_{sea}(p)}{u(\pi)\,u(p) + q_{sea}(\pi)\,d(p)} \, < \, 
\frac{4 u(\pi)\,q_{sea}(p)}{d(\pi)\,d(p) + 8 q_{sea}(\pi)\,u(p)} \; , 
\label{eq:csivI-test2-d}
\ee
namely $c_{Siv\, \pi^-}^I$ is closer to $c_{Siv}^{II}=1$ than $c_{Siv\, \pi^+}^I$. 
This fact is responsible for the plot in Fig.~\ref{fig:pi+sealowM} showing much 
more scattered events than in Fig.~\ref{fig:pi-sealowM}. Given the lowest $M$ range 
here explored, the situation displayed in Fig.~\ref{fig:pi+sealowM} represents a sort 
of ideal upper limit to the potential discrepancy between the two methods for 
calculating the Drell-Yan SSA.

\begin{figure}[h]
\centering
\includegraphics[width=7cm]{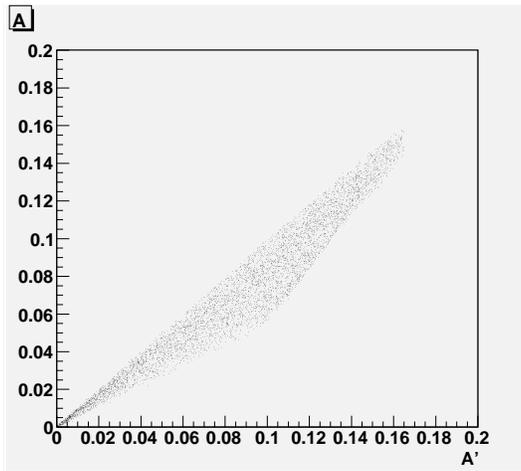}
\caption{Scatter plot in the same conditions as Fig.~\protect{\ref{fig:pi-sealowM}}, 
but using the parametrization of the Sivers function from 
Ref.~\protect{\cite{Anselmino:2005ea}}.}
\label{fig:pi-TO}
\end{figure}

We now turn to the comparison between "framework I" and "framework II" with realistic
parametrizations of the Sivers function and including the "sea dilution effect". In
Fig.~\ref{fig:pi-TO}, the scatter plot of $A$ versus $A'$ is shown for the 
$\pi^- p^\uparrow \to \mu^+ \mu^- X$ process with $1.5<M<2.5$ GeV using the Sivers
function of Ref.~\cite{Anselmino:2005ea}, i.e. using $c_{Siv}^A$ from
Eqs.~(\ref{eq:csiv-A}) and (\ref{eq:csiv-Abis}). The discrepancy is not large, if we
consider the potentially dangerous low $M$ range. 

\begin{figure}[h]
\centering
\includegraphics[width=7cm]{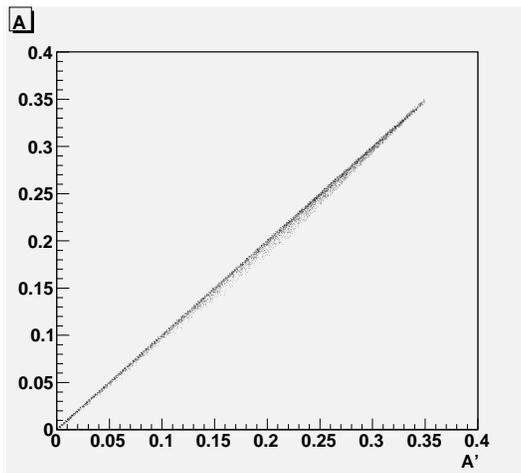}
\caption{Scatter plot in the same conditions as previous figure, 
but using the parametrization of the Sivers function from 
Ref.~\protect{\cite{Bianconi:2005yj,Bianconi:2006hc}} and for muon invariant mass
$4<M<6$ GeV.}
\label{fig:pi-noi}
\end{figure}

In Fig.~\ref{fig:pi-noi}, the same situation of the previous case is simulated at
$4<M<6$ GeV using the Sivers function from 
Ref.~\cite{Bianconi:2005yj,Bianconi:2006hc}. Here, the agreement between the two
methods is much more evident also because the higher $M$ range considered corresponds
to a safer valence domain in $x_2$, approximately around 0.2-0.4. 

\begin{figure}[h]
\centering
\includegraphics[width=7cm]{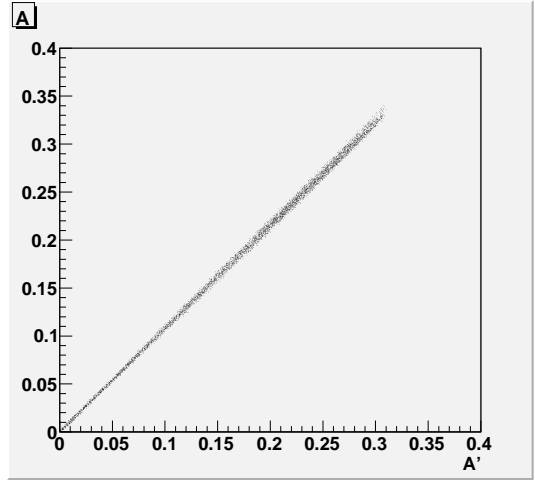}
\caption{Scatter plot in the same conditions as previous figure, 
but for the $\bar{p} p^\uparrow \to \mu^+ \mu^- X$ process.}
\label{fig:antip-noi}
\end{figure}

Next, in Fig.~\ref{fig:antip-noi} we reconsider the situation of the previous figure
but for the $\bar{p} p^\uparrow \to \mu^+ \mu^- X$ process. At the level of leading
valence contributions to the asymmetry, the $\pi^- p^\uparrow$ and the 
$\bar{p} p^\uparrow$ collisions are equivalent because they are both dominated by the
$\bar{u} u^\uparrow$ annihilation. Hence, the persisting close similarity of $A$ and 
$A'$ indicates that the main origin of the discrepancy between "framework I" 
and "framework II" comes from the "sea dilution effect", namely from including 
or neglecting the contribution of unpolarized sea (anti)quarks in the 
denominator of the SSA.

\begin{figure}[h]
\centering
\includegraphics[width=7cm]{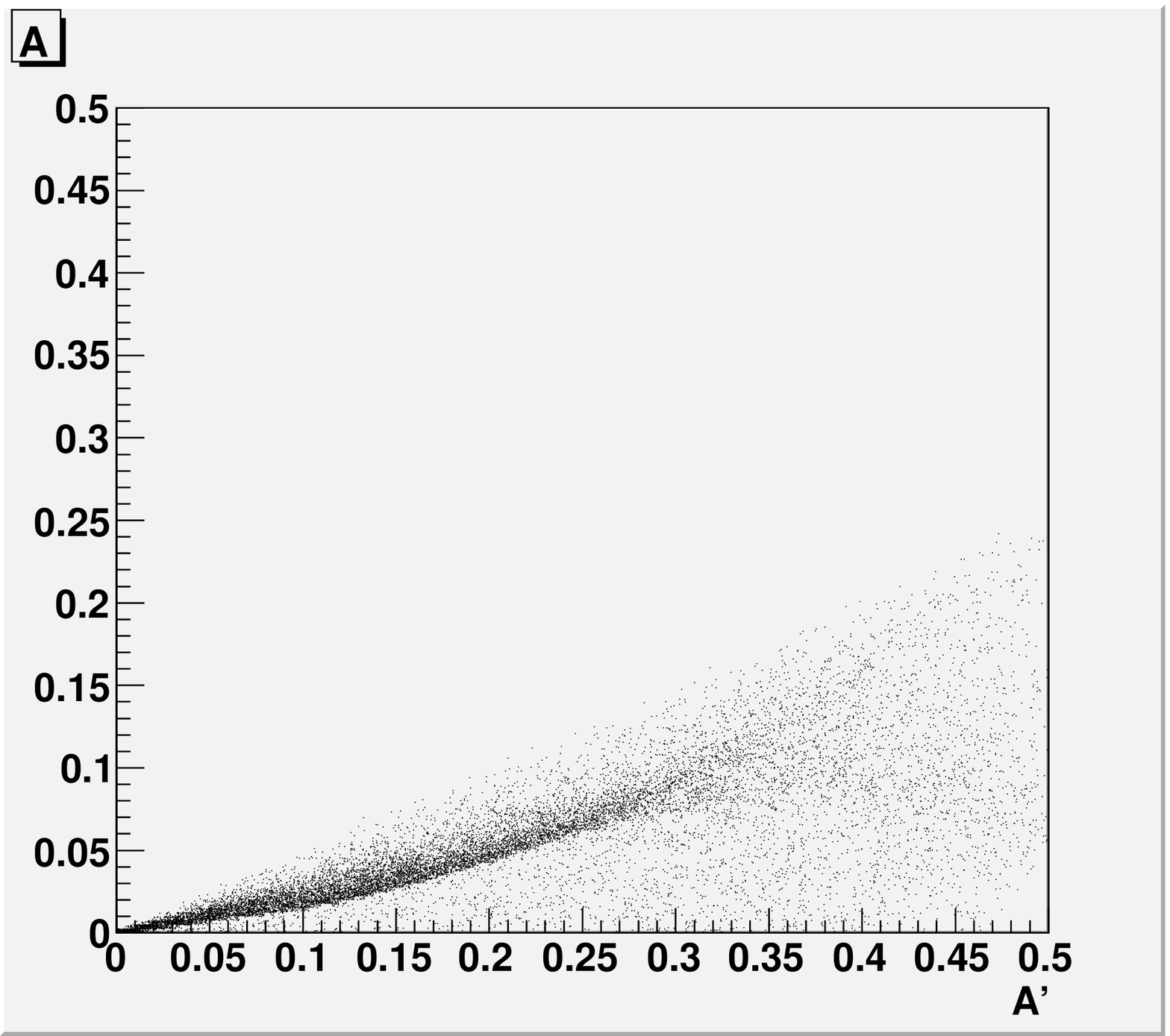}
\caption{Scatter plot in the same conditions as Fig.~\protect{\ref{fig:pi-TO}}, 
but for the $\pi^+ p^\uparrow \to \mu^+ \mu^- X$ process.}
\label{fig:pi+TO}
\end{figure}

Last, in Fig.~\ref{fig:pi+TO} we show the scatter plot of $A$ versus $A'$ in the same
conditions as in Fig.~\ref{fig:pi-TO} but for the 
$\pi^+ p^\uparrow \to \mu^+ \mu^- X$ process. The large discrepancies confirm the
findings about Eq.~(\ref{eq:csivI-test2-d}) that were displayed in 
Fig.~\ref{fig:pi+sealowM} using a simpler naive parametrization of 
the Sivers function: with the 
$\pi^+$ probe the "sea dilution effect" is emphasized and the approximations 
introduced in "framework II" with Eq.~(\ref{eq:csiv-Abis}) are somewhat not justified. 
The effect is also emphasized by the low $M$ range considered.

In summary, when the spin asymmetries $A$ and $A'$ are simulated using realistic
parametrizations of the Sivers function, which are all based on the dominance of the
transversely polarized $u^\uparrow$ valence distribution in the transversely polarized
$p^\uparrow$ parent hadron, their results are very close provided that the effect of
the unpolarized sea (anti)quarks can be neglected. This condition is fulfilled by the
$\pi^-$ and $\bar{p}$ probes at not too low muon invariant masses $M$, which
correspond to the safe valence domain in $x$. In these cases, the originating 
schemes "framework I" and "framework II" can be considered equivalent. 

However, the last statement can be misleading, because the realistic parametrizations
of the Sivers function are obtained in the "framework II", namely by
building the SSA by 
neglecting the unpolarized sea (anti)quarks. It means that any contribution from the 
sea, which is intrinsically contained in the experimental measurement of the 
asymmetry, is effectively reproduced in the parametrization, particularly in the 
behaviour at low $x$. For example, this "hidden sea effect" is contained in the 
$\alpha_q$ parameter of the $x^{\alpha_q}$ factor in Eq.~(\ref{eq:csiv-A}), which is 
determined from a fit of experimental SSA based on valence $q=u,d$  
degrees of freedom only~\cite{Anselmino:2005ea}. Hence, it can be over- or 
under-estimated when the related Sivers function is plugged into expressions of SSA 
constructed with "framework I" or "framework II". 

Actually, the emerging issue here is that "framework I", which looks more correct
because it properly includes the unpolarized sea contribution in the denominator of
the SSA, could produce a sort of double counting when employing a Sivers
function in the numerator that is parametrized with valence quarks only.
Consequently, it is not obvious that the approximations adopted in "framework II" (and
systematically used in the analysis of several Drell-Yan SSA in 
Refs.~\cite{Bianconi:2004wu,Bianconi:2005bd,Bianconi:2005yj,Bianconi:2006hc}) are
crude ones; rather, they look like the most appropriate framework for using the
parametrized Sivers functions available in the literature. However, this last
statement should be generalized with some care. In fact, the large discrepancies
observed in the results obtained with "framework I" and "framework II" using $\pi^+$ probes,
suggest that each physics case should be separately considered.

Finally, as anticipated in Sec.~\ref{sec:formulae}, we reconsider the $\nu$ term in
Eq.~(\ref{eq:mcS}), which is most likely related to the violation of the Lam-Tung sum
rule~\cite{Falciano:1986wk,Guanziroli:1987rp,Conway:1989fs}. The corresponding 
azimuthal $\cos 2\phi$ asymmetry in the unpolarized Drell-Yan cross section was 
simulated in Ref.~\cite{Bianconi:2004wu} using the simple parametrization of 
Ref.~\cite{Boer:1999mm} and testing it against the previous measurements of 
Refs.~\cite{Falciano:1986wk,Guanziroli:1987rp,Conway:1989fs}. The Boer-Mulders function
$h_1^\perp(x)$ was parametrized in Ref.~\cite{Boer:1999mm} exactly in the 
"framework II" but with no flavor dependence, because of few available data at that 
time. The latter span only the region $x>0.3$, while for $x<0.3$ the $\nu$ function was 
assumed to be independent from $x$. Hence, the present parametrization of $h_1^\perp$ 
does not effectively include any "sea dilution effect". The situation could be improved 
by either refitting the Drell-Yan data of 
Refs.~\cite{Falciano:1986wk,Guanziroli:1987rp,Conway:1989fs} using "framework I" and 
an $x$-independent $\nu$ function for $x<0.3$, or using "framework II" and introducing a
specific $x^\alpha$ power law to extrapolate the behaviour at low $x$. Moreover, new data
have been recently released for Drell-Yan muon pairs produced in high-energy
proton-deuteron collisions~\cite{Zhu:2006gx}, that show no evidence of a $\cos 2\phi$
asymmetry at very low $x$. In any case, the few available data do not allow yet a 
full flavor-dependent analysis and any conclusion is, consequently, premature.

\section{Conclusions}
\label{sec:end}

In this paper, we performed numerical simulations of the socalled Sivers 
effect~\cite{Sivers:1990cc}, as it can be isolated in single spin asymmetries (SSA) of 
the distribution of Drell-Yan muon pairs produced from collisions of transversely 
polarized protons and different hadronic probes. Several measurements of such SSA are 
planned by experimental collaborations (RHIC at BNL, COMPASS at CERN, PANDA and PAX at 
GSI, and, possibly, also future experiments at JPARC). The goal is the extraction of 
the socalled Sivers function $f_{1\sT}^\perp$, a leading-twist parton distribution that 
can give direct insight into the orbital motion and the spatial distribution of hidden 
confined partons, with interesting connections with the problem of the proton spin sum 
rule and the powerful formalism of Generalized Parton 
Distributions~\cite{Burkardt:2003je}. In particular, it 
would be extremely important to verify the peculiar universality properties of such 
parton density, namely its predicted sign change with respect to the same 
$f_{1\sT}^\perp$ as it is parametrized to reproduce the SSA in SIDIS 
measurements~\cite{Bianconi:2005yj,Bianconi:2006hc}. This theorem is based on very 
general assumptions and it represents a formidable test of QCD 
universality~\cite{Collins:2002kn}. 

One of the main features of the SIDIS phenomenological parametrizations is the 
approximation of neglecting both polarized and unpolarized nonvalence partons (see, 
e.g., Ref.~\cite{Anselmino:2005ea}), which actually amounts to effectively include 
their contribution in the fitting parameters of the valence partons for the $x$ range
considered. We conventionally name this the "hidden sea effect". In a series of previous 
papers~\cite{Bianconi:2004wu,Bianconi:2005bd,Bianconi:2005yj,Bianconi:2006hc}, we 
performed numerical simulations of Drell-Yan SSA with transversely polarized protons 
using colliding protons, antiprotons, and pions, in various kinematics of interest 
for the planned experiments. In our Monte Carlo code, the Sivers effect was consistently
simulated within the same approach, but we further conveniently made a suitable
flavor average of the valence contribution which allows for a great simplification
of formulae. We conventionally name this scheme as "framework II". Here, we consider
also the socalled "framework I", where we release the approximation about the flavor
average and we include also the unpolarized nonvalence contribution, which shows up
mainly in the denominator of the SSA; as such, we conventionally
refer to the "sea dilution effect".

We have explored the deviations of "framework II" from the more appropriate "framework
I" in the simulation of the Sivers effect for the $H p^\uparrow \to \mu^+ \mu^- X$
process at $s=100$ GeV$^2$ with $H=\bar{p}, \pi^-, \pi^+$ and different muon invariant
masses. It turns out that the two approaches become approximately equivalent when the 
polarized distributions weakly depend on flavor (or, equivalently, one flavor dominates 
the flavor sum in the numerator of the SSA), and for sufficiently high muon invariant 
masses (typically, $4<M<6$ GeV), which correspond to $\langle x \rangle$ in the safe 
valence domain where the contribution of sea (anti)quarks becomes negligible 
(assuming that also transversely polarized sea (anti)quarks can be neglected, as well). 
These conditions can be fulfilled when using the $\pi^-$ and $\bar{p}$ probes; on the
contrary, the valence structure of $\pi^+$ emphasizes the role of the "sea dilution
effect" and leads to larger discrepancies. 

However, we must remark that the parametrizations of the Sivers function are basically 
obtained using "framework II", since all sea (anti)quarks are neglected. Hence, the
"hidden sea effect" contained in the values of the parameters can be over- or 
under-estimated when the related Sivers function is plugged into expressions of SSA 
constructed with "framework I" or "framework II". Actually, the "framework I" appears
even less appropriate, since it could lead to a double counting of the contribution of
unpolarized sea (anti)quarks; in our jargon, the "sea dilution effect" induced by the
denominator of the SSA would describe the same mechanisms as the "hidden sea effect" 
contained in the fitting parameters of the valence quark distributions. Correspondingly,
the approximations adopted in "framework II" (and systematically used in the analysis 
of Refs.~\cite{Bianconi:2004wu,Bianconi:2005bd,Bianconi:2005yj,Bianconi:2006hc}) look 
like the most appropriate approach for using the parametrized Sivers functions presently
available in the literature. However, this statement should be taken with some care,
because there are cases like the $\pi^+ p^\uparrow$ collisions where the "framework II"
largely deviates from "framework I" in any kinematics and seems not well justified. 

In our work, the Sivers effect has been used as a test case since the abundance of
data allows for a flavor-dependent analysis. In principle, the arguments can 
be generalized to other interesting azimuthal asymmetries in Drell-Yan processes, such 
as the Boer-Mulders effect or the violation of the Lam-Tung sum rule. But in the former
case, there are no experimental data, while in the latter the few ones 
available~\cite{Falciano:1986wk,Guanziroli:1987rp,Conway:1989fs,Zhu:2006gx} do not 
permit to discriminate the contributions of each flavor and prevent from coming to 
definite conclusions.


\begin{acknowledgments}

This work is part of the European Integrated Infrastructure Initiative in Hadron
Physics project under the contract number RII3-CT-2004-506078.

\end{acknowledgments}


\bibliographystyle{apsrev}
\bibliography{hadron.bib}

\end{document}